\documentclass[final,3p,times,twocolumn]{elsarticle}
\biboptions{sort&compress}

\usepackage{amssymb,mathrsfs}
\usepackage{amsthm,amsmath}
\usepackage{bm}
\usepackage{multirow,booktabs}
\usepackage{slashed}
\usepackage{threeparttable}

\usepackage[
bookmarks=true,colorlinks,%
citecolor=blue,linkcolor=blue,
breaklinks=true]{hyperref}

\begin{document}
	
	
	\hyphenpenalty=6000
	\tolerance=3000
	
	\begin{frontmatter}
		
		
		
        \title{Constraints on $\Lambda N$ Effective Interactions from Mirror Hypernuclei in a Deformed Relativistic Hartree-Bogoliubov Model}
		
		
		\author[GXNU,GXNPT] {Yu-Ting Rong  \corref{cor1}}
        \ead{rongyuting@gxnu.edu.cn}
        \cortext[cor1]{Corresponding author.}
        
        \author[GXNU,GXNPT] {Dan Yang }
        
        \author[YZU] {Cheng-Jun Xia \corref{cor1}}
        \ead{cjxia@yzu.edu.cn}
        
        \author[ZZU] {Ting-Ting Sun \corref{cor1}}
        \ead{ttsunphy@zzu.edu.cn}

		\affiliation[GXNU]{%
		    organization={Department of Physics, Guangxi Normal University},%
		    city={Guilin},%
		    postcode={541004},%
		    country={China}}	
	    
	    \affiliation[GXNPT]{%
	    	organization={Guangxi Key Laboratory of Nuclear Physics and Technology, Guangxi Normal University},%
	    	city={Guilin},%
	    	postcode={541004},%
	    	country={China}}
		
		\affiliation[YZU]{%
    	organization={Center for Gravitation and Cosmology, College of Physical Science and Technology, Yangzhou University},%
    	city={Yangzhou},%
    	postcode={225009},
    	country={China}}

        \affiliation[ZZU]{%
    	organization={School of Physics, Zhengzhou University},%
    	city={Zhengzhou},%
    	postcode={450001},%
    	country={China}}

		\begin{abstract}
			We investigate the ground-state properties of four mirror hypernuclei pairs--$^{10}_\Lambda$Be-$^{10}_\Lambda$B, $^{12}_\Lambda$B-$^{12}_\Lambda$C,  $^{16}_\Lambda$N-$^{16}_\Lambda$O, and $^{40}_\Lambda$K-$^{40}_\Lambda$Ca--within the deformed relativistic Hartree-Bogoliubov framework, analyzing their connection to $\Lambda N$ effective interactions. Systematic calculations with eight distinct effective interactions reveal linear correlations between mirror hypernuclei in $\Lambda$ separation energies and charge radii. The charge symmetry breaking effects, quantified through $\Lambda$ separation energy differences, exhibit a positive correlation with the SU(3) flavor symmetry violation. We emphasize that constraints derived from $A=10$ and $A=12$ hypernuclear pairs must explicitly incorporate rotational energy correction effects. Precision measurements of the (near) spherical $A=16$ and $A=40$ mirror systems are proposed as critical benchmarks for refining the isospin part of the hyperon-nucleon interactions.
	
		\end{abstract}

		
	\end{frontmatter}
	
\section{Introduction}\label{sec1}

Hyperon-nucleon ($YN$) and hyperon-hyperon ($YY$) interactions are fundamental constituents of baryon-baryon interactions. 
The precise characterization of these interactions is essential for understanding hypernuclear structure and the equation of state governing compact stars~\cite{Gal2016_RMP88-035004,Tolos2020_PPNP112-103770}.
Despite significant progress in $YN$ and $YY$ interaction studies through $e^+e^-$ collision experiments~\cite{Ablikim2024_PRC109-L052201}, state-of-the-art lattice QCD simulations~\cite{Beane2007_NPA794-62,Nemura2009_PLB673-136,Doi2018_EPJWoC175-05009,Nemura2018_EPJWoC175-05030}, and developments in chiral effective field theory~\cite{Polinder2006_NPA779-244,Haidenbauer2013_NPA915-24,Kohno2019_PRC100-024313,Ren2020_PRC101-034001,Song2022_PRC105-035203,Liu2023_CPC47-024108}, the direct determination of bare $YN$ and $YY$ interactions remains experimentally difficult due to hyperons' inherently short lifetimes. This experimental limitation necessitates the development of in-medium $YN$ and $YY$ interaction models derived from SU(3) flavor symmetry constraints combined with hypernuclear structure data~\cite{Zhang1982_PLB108-261,Levai1998_PLB433-250,Rijken2010_PTPSupp185-14,van-Dalen2014_PLB734-383,Schulze2014_PRC90-047301,Nagels2019_PRC99-044003,Rong2020_PLB807-135533,Rong2021_PRC104-054321,Ding2022_PRC106-054311}. 
Quantitative constraints on these interactions emerge from systematic studies of single-$\Lambda$, double-$\Lambda$~\cite{Takahashi2001_PRL87-212502},
$\Xi$ hypernuclei~\cite{Hayakawa2021_PRL126-062501}, and antihypernuclei~\cite{STAR2020_NatPhys16-409,STAR2024_Nature632-1026}, which collectively provide critical benchmarks for theoretical models.


Mirror nuclei, defined as pairs of isobaric nuclei sharing identical mass number $A$ while exhibiting interchanged proton number $Z$ and neutron number $N$, serve as critical tools for probing the isospin-dependent component of nucleon-nucleon ($NN$) interactions in conventional nuclear systems. Notably, comparative studies of charge radius disparities in mirror nuclei have been established as a sensitive methodology to constrain the slope parameter of nuclear symmetry energy~\cite{Wang2013_PRC88-011301,Brown2017_PRL119-122502,Yang2018_PRC97-014314,Brown2020_PRR2-022035}. Precision mass relations derived from mirror nuclear systems enable robust predictions for proton-rich nuclei masses~\cite{Tian2013_PRC87-014313,Bao2016_PRC94-044323,Zong2019_PRC100-054315}.
In the hypernuclear domain, seven experimentally confirmed mirror hypernuclear pairs have been identified: $^{4}_\Lambda$H-$^{4}_\Lambda$He, $^{7}_\Lambda$Li-$^{7}_\Lambda$Be, $^{8}_\Lambda$Li-$^{8}_\Lambda$Be, $^{9}_\Lambda$Li-$^{9}_\Lambda$B, $^{10}_\Lambda$Be-$^{10}_\Lambda$B, $^{12}_\Lambda$B-$^{12}_\Lambda$C, and $^{16}_\Lambda$N-$^{16}_\Lambda$O~\cite{Hashimoto2006_PPNP57-564,Gal2016_RMP88-035004}. 
These systems provide crucial insights into 
$\Lambda N$ interaction dynamics, particularly regarding charge symmetry breaking (CSB) phenomena in hypernuclei~\cite{Hiyama2009_PRC80-054321,Nogga2013_NPA914-140,Gal2015_PLB744-352,Yamamoto2015_PRL115-222501,Schaefer2022_PRC106-L031001}. The analysis of binding energy differences and spectral properties across these mirror pairs offers unique constraints on the isospin structure of $YN$ forces.

In our previous investigation utilizing the relativistic mean-field (RMF) framework, we have demonstrated that the $\Lambda$ separation energies of two mirror hypernuclear pairs, namely 
$^{12}_\Lambda$B-$^{12}_\Lambda$C and $^{16}_\Lambda$N-$^{16}_\Lambda$O, could not be adequately described through conventional parametrizations incorporating density-dependent scalar, vector, and tensor potentials ~\cite{Rong2021_PRC104-054321}. 
We have also found that center-of-mass effect and rotational symmetry restoration are not negligible for describing the ground-state properties of light nuclear systems~\cite{Rong2023_PRC108-054314}. 
Besides, several mirror hypernuclei under investigation--along with their core nuclei-- exhibit intrinsic deformation.
Considering that the observed mirror hypernuclei are in light mass region, the above findings underscore the necessity for employing a self-consistent deformed nuclear structure model that incorporates both center-of-mass and rotational corrections to study mirror hypernuclear pairs. Such an advancement would enable more rigorous constraints on the in-medium $\Lambda N$ effective interactions, particularly those governing CSB phenomena in hypernuclear systems.

Therefore, the objective of this work is to investigate the properties of mirror hypernuclei through a deformed mean-field framework that incorporates nuclear deformation, microscopic center-of-mass corrections, and rotational energy corrections. Our aim is to identify novel experimental observables capable of constraining the isospin-dependent component of the $\Lambda N$ effective interaction.
The manuscript is structured as follows: Section \ref{sec:model} presents the theoretical formalism of the multidimensionally constrained relativistic Hartree-Bogoliubov (MDC-RHB) model extended for hypernuclear systems. In Section \ref{sec:results}, we perform a comprehensive analysis of bulk properties, including comparisons of binding energy differences, charge radius variations, and deformation parameters between four selected mirror hypernuclear pairs and their core nuclei. These calculations are conducted using eight different effective interactions to assess model dependence. Finally, Section \ref{sec:summary} synthesizes our findings and discusses their implications for constraining $\Lambda N$ interaction.

\section{Theoretical framework}\label{sec:model}

The RHB theory provides a unified description of the mean field and the pairing correlations via the Bogoliubov transformation. The key point of this theory is to solve the RHB equation consisting of single-particle Hamiltonian $h_B$ for a baryon (neutron, proton or hyperon) and pairing field $\Delta$, i.e.,
\begin{equation}
	\label{eq:rhb}
	\int d^{3}\bm{r}^{\prime}
	\left( \begin{array}{cc} h_B-\lambda  &  \Delta                      \\ 
		-\Delta^{*}   & -h_B+\lambda \end{array} 
	\right)
	\left( \begin{array}{c} U_{k} \\ V_{k} \end{array} \right)
	= E_{k}
	\left( \begin{array}{c} U_{k} \\ V_{k} \end{array} \right),
\end{equation} 
where $\lambda$ is the Fermi energy, $E_k$ is the quasiparticle energy, 
and $(U_k,V_k)^T$ is the quasiparticle wave function, respectively. The Hamiltonian for nucleons has been well-established~\cite{Serot1986_ANP16-1,Reinhard1989_RPP52-439,Ring1996_PPNP37-193,
	Bender2003_RMP75-121,Vretenar2005_PR409-101,Meng2006_PPNP57-470,
	Niksic2011_PPNP66-519,Liang2015_PR570-1,Meng2015_JPG42-093101,
	Zhou2016_PS91-063008} and that for $\Lambda$ hyperon is~\cite{Schaffner1994_AP235-35} 
\begin{equation}
	h_\Lambda(\bm{r})= \bm{\alpha} \cdot \bm{p} + V_\Lambda(\bm{r}) + 
	\beta \left( m_\Lambda + S_\Lambda(\bm{r}) \right) + 
	T_\Lambda(\bm{r}),
\end{equation} 
where $m_\Lambda$ is the mass of $\Lambda$, and $S_\Lambda(\bm{r}) = g_{\sigma\Lambda} \sigma  (\bm{r})$ is the scalar potential, $V_\Lambda(\bm{r}) = g_{\omega\Lambda} \omega_0(\bm{r})$ is the vector potential, and $T_\Lambda(\bm{r}) =-\frac{f_{\omega\Lambda\Lambda}}{2m_\Lambda} 
\beta (\bm{\alpha} \cdot \bm{p}) \omega_0(\bm{r})$ is the  tensor potential.
$g_{\sigma\Lambda}$, $g_{\omega\Lambda}$, and $f_{\omega\Lambda\Lambda}$ are coupling constants for $\sigma$ field and time-like component of $\omega$ meson field. 
The pairing potential reads
\begin{equation}
	\begin{aligned}
		\Delta(\bm{r}_{1}\sigma_{1},\bm{r}_{2}\sigma_{2})
		&=  \int d^{3}\bm{r}_{1}^{\prime} d^{3}\bm{r}_{2}^{\prime} 
		\sum_{\sigma_{1}^{\prime}\sigma_{2}^{\prime}} \\
		&V(\bm{r}_{1}         \sigma_{1},          \bm{r}_{2}         \sigma_{2},
		\bm{r}_{1}^{\prime}\sigma_{1}^{\prime}, \bm{r}_{2}^{\prime}\sigma_{2}^{\prime}) 
		\kappa 
		(\bm{r}_{1}^{\prime}\sigma_{1}^{\prime}, 
		\bm{r}_{2}^{\prime}\sigma_{2}^{\prime}),
	\end{aligned}
\end{equation}
where 
$V$ is the effective pairing interaction and $\kappa$ is the pairing tensor. In this work, a separable pairing force of finite range~\cite{Tian2009_PLB676-44,Tian2009_PRC80-024313} is adopted for the paired nucleons, and the unpaired baryon is blocked to a certain single-particle level.

Under the mean field and no-sea approximation, the RHB equation (\ref{eq:rhb}) can be solved in different bases~\cite{Horowitz1981_NPA368-503,Gambhir1990_APNY198-132,Zhou2003_PRC68-034323,Typel2018_FPhys6-73} or 3D lattice~\cite{Ren2017_PRC95-024313}. In the present work, we use the MDC-RHB model~\cite{Zhou2016_PS91-063008,Zhao2017_PRC95-014320} in which an axially deformed harmonic oscillator (ADHO) basis~\cite{Gambhir1990_APNY198-132} is used. The deformation parameter $\beta_{\lambda \mu}$ is calculated by
\begin{equation}
	\beta_{\lambda\mu}=\dfrac{4\pi}{3A R^\lambda} Q_{\lambda\mu},
\end{equation}
where the intrinsic multipole moment $Q_{\lambda\mu}=\int d^3 r \rho^\upsilon(r) r^\lambda Y_{\lambda\mu}(\Omega)$, $Y_{\lambda\mu}$ the spherical harmonics, and $R=1.2A^{1/3}$ fm. In the MDC-RHB model, thanks to the usage of ADHO basis, the potentials and densities are invariant under the $V_4$ symmetry, i.e., all $\beta_{\lambda\mu}$ with even $\mu$ can be considered simultaneously.

Both the center-of-mass and rotational corrections are sizable in describing the ground-state properties of deformed light nuclei~\cite{Rong2023_PRC108-054314}. Particularly for the cases where the mass number $A$ of the observed mirror hypernuclei are less than 16. Therefore, binding energy $E_{\rm B}$ for a single-$\Lambda$ hypernucleus is comprised of mean-field energy $E_{\rm MF}$, center-of-mass energy $E_{\rm c.m.}$, and rotational energy $E_{\rm rot.}$ as follows
\begin{equation}
	E_{\rm B}=-E_{\rm MF}-E_{\rm c.m.}-E_{\rm rot.}.
\end{equation}
The center-of-mass energy is calculated by
\begin{equation}\label{eq:E_mic}
	E_{\rm c.m.}^{\rm mic}=-\dfrac{\langle P_{\rm c.m.}^2\rangle }{2M},
\end{equation}
where $M=Nm_n+Zm_p+m_\Lambda$, and 
\begin{equation}\label{eq:cm-P-square}	
	\begin{aligned}
		\langle P_{\rm c.m.}^2\rangle=&
		\sum_i \upsilon_i^2 p_{ii}^2-\sum_{i,j}\upsilon_i^2\upsilon_j^2 p_{ij}p_{ij}^* 
		+\sum_{i,j}\upsilon_iu_i\upsilon_ju_j p_{ij}p_{\bar{i}\bar{j}}.
	\end{aligned}
\end{equation}
In eq.~(\ref{eq:cm-P-square}), $i$ and $j$ denote the quasiparticle states, $\upsilon_i$ is the occupation probability, and $u_i^2+\upsilon_i^2=1$.
The rotational energy is calculated by
\begin{equation}\label{eq:Erot}
	E_{\rm rot.}=-\dfrac{1}{2} \sum_{k=1}^3 \dfrac{\langle J_k^2\rangle}{I_k},
\end{equation}
where $k$ denotes the axis of rotation, $J_k$ the component of the angular momentum
in the body-fixed frame of a nucleus. The 
moment of inertia $I_k$ is a linear combination of Inglis-Belyaev formula and the moment of inertia of rigid rotor, i.e., $I_k=0.8I_k^{\rm IB}+0.2I_k^{\rm rigid}$, with the Inglis-Belyaev formula~\cite{Inglis1956_PR103-1786,Beliaev1961_NP24-322}
\begin{equation}
	I_k^{\rm IB}=\sum_{i,j} \dfrac{(u_i\upsilon_j-\upsilon_i u_j)^2}{E_i+E_j} |\langle i|J_k|j\rangle|^2.
\end{equation}
The charge radius $r_{\rm ch}$ is obtained by~\cite{Sugahara1994_NPA579-557}
\begin{equation}\label{eq:rc_from_rp}
	r_{\rm ch}^2=r_p^2+(0.862~{\rm fm})^2-(0.336~{\rm fm})^2 N/Z,
\end{equation}
where $r_p$ is the charge distributions of point-like proton.

\begin{table*}[htb!]
	\caption{Ground-state properties of the selected mirror nuclear pairs ($^{9}$Be-$^{9}$B, $^{11}$B-$^{11}$C, $^{15}$N-$^{15}$O, and $^{39}$K-$^{39}$Ca) calculated using four distinct effective interactions: PK1~\cite{Long2004_PRC69-034319}, NLSH~\cite{Sharma1993_PLB312-377}, DD-ME2~\cite{Lalazissis2005_PRC71-024312}, and PKDD~\cite{Long2004_PRC69-034319}. }\label{tab:core-nuclei} 
	\doublerulesep 0.1pt \tabcolsep 2.5pt
	\begin{tabular}{ccccccccc}
		\hline
		\hline
		Force & $E$ (MeV) & $r_{\rm ch}$ (fm)
		& $\beta_{20}$& $E$ (MeV)  & $r_{\rm ch}$ (fm)
		&$\beta_{20}$&$\Delta E$ (MeV)
		& $\Delta r_{\rm ch}$ (fm)\\
		\hline           
		{} & \multicolumn{3}{c}{$^{9}$Be}     &  \multicolumn{3}{c}{$^{9}$B}   \\
		\cmidrule(lrr){2-4}  \cmidrule(lrr){5-7}  
		PK1   &  55.843 & 2.389  & $-$0.340   & 53.348    & 2.578  & $-$0.348  & 2.495  & $-$0.189 \\
		NLSH  &  59.016 & 2.373  & $-$0.327   & 56.616    & 2.555  & $-$0.334  & 2.400  & $-$0.182 \\
		DD-ME2&  55.239 & 2.464  & $-$0.361   & 52.772    & 2.651  & $-$0.371  & 2.467  & $-$0.187 \\
		PKDD  &  55.171 & 2.417  & $-$0.345   & 52.695    & 2.604  & $-$0.353  & 2.476  & $-$0.187 \\
		Exp.
		&  58.164~\cite{Wang2021_ChinPhysC45-030003} 
		&2.5190(120)~\cite{Angeli2013_ADNDT99-69}
		&            & 56.314~\cite{Wang2021_ChinPhysC45-030003}
		&        &           &  1.85  &          \\
		\hline
		{}    & \multicolumn{3}{c}{$^{11}$B}  &  \multicolumn{3}{c}{$^{11}$C}   \\
		\cmidrule(lrr){2-4}  \cmidrule(lrr){5-7}  
		PK1   &  78.800 & 2.440  & $-$0.324   & 75.814    & 2.559  & $-$0.331  & 2.986  & $-$0.119 \\
		NLSH  &  80.205 & 2.430  & $-$0.304   & 77.320    & 2.544  & $-$0.312  & 2.885  & $-$0.114 \\
		DD-ME2&  77.996 & 2.506  & $-$0.350   & 74.953    & 2.616  & $-$0.358  & 3.043  & $-$0.110 \\
		PKDD  &  78.190 & 2.472  & $-$0.342   & 75.139    & 2.587  & $-$0.348  & 3.051  & $-$0.115 \\
		Exp.  &  76.205~\cite{Wang2021_ChinPhysC45-030003}   
		& 2.4060(294)~\cite{Angeli2013_ADNDT99-69}
		&            &73.441~\cite{Wang2021_ChinPhysC45-030003}     
		&  2.32(11)~\cite{Zhao2024_PLB858-139082}       
		&           & 2.764  &  0.09    \\
		\hline
		{}    & \multicolumn{3}{c}{$^{15}$N}  &  \multicolumn{3}{c}{$^{15}$O}  &      \\
		\cmidrule(lrr){2-4}  \cmidrule(lrr){5-7}  
		PK1   & 116.397 & 2.613  & 0.000      &  112.665  & 2.707  & 0.000     & 3.732  & $-0.094$ \\
		NLSH  & 117.196 & 2.614  & 0.000      &  113.564  & 2.705  & 0.000     & 3.632  & $-0.091$ \\
		DD-ME2& 116.284 & 2.652  & 0.000      &  112.539  & 2.735  & 0.000     & 3.745  & $-0.083$ \\
		PKDD  & 116.060 & 2.621  & 0.000      &  112.294  & 2.712  & 0.000     & 3.766  & $-0.091$ \\
		Exp.  & 115.492~\cite{Wang2021_ChinPhysC45-030003}
		& 2.6058(80)~\cite{Angeli2013_ADNDT99-69} 
		&            &  111.955~\cite{Wang2021_ChinPhysC45-030003}  &        &           & 3.537  &           \\
		\hline
		{}    & \multicolumn{3}{c}{$^{39}$K}  &  \multicolumn{3}{c}{$^{39}$Ca}   \\
		\cmidrule(lrr){2-4}  \cmidrule(lrr){5-7}  
		PK1   &333.504  & 3.402  & $-$0.038   & 326.201   & 3.444  & $-$0.039  & 7.303  & $-$0.042 \\
		NLSH  & 331.354 & 3.408  & $-$0.037   & 324.139   & 3.449  & $-$0.036  & 7.215  & $-$0.041 \\
		DD-ME2& 333.884 & 3.427  & $-$0.028   & 326.473   & 3.461  & $-$0.029  & 7.411  & $-$0.034 \\
		PKDD  & 333.247 & 3.403  & $-$0.034   & 325.856   & 3.442  & $-$0.035  & 7.391  & $-$0.039 \\
		Exp.
		& 333.724~\cite{Wang2021_ChinPhysC45-030003} 
		& 3.4349(19)~\cite{Angeli2013_ADNDT99-69}
		&            & 326.417~\cite{Wang2021_ChinPhysC45-030003}   
		& 3.4595(25)~\cite{Angeli2013_ADNDT99-69}
		&           & 7.307  & $-$0.0246  \\
		\hline         
		\hline	
	\end{tabular}
\end{table*}


\begin{table*}[htb!]
	\caption{Ground-state properties of the selected mirror hypernuclear pairs ($^{10}_\Lambda$Be-$^{10}_\Lambda$B, $^{12}_\Lambda$B-$^{12}_\Lambda$C, $^{16}_\Lambda$N-$^{16}_\Lambda$O, and $^{40}_\Lambda$K-$^{40}_\Lambda$Ca) calculated using eight distinct effective interactions: PK1-Y1~\cite{Wang2013_CTP60-479}, NLSH-A~\cite{Mares1994_PRC49-2472}, DD-ME2-Y$i$ ($i=1,~2,~3$), and PKDD-Y$i$ ($i=1,~2,~3$)~\cite{Rong2021_PRC104-054321}.}\label{tab:hypernuclei-gs} 
	\doublerulesep 0.1pt \tabcolsep 4pt
	\begin{tabular}{ccccccccc}
		\hline
		\hline
		Force          & $S_\Lambda$ (MeV) & $r_{\rm ch}$ (fm) & $\beta_{20}$     & $S_\Lambda$ (MeV) & $r_{\rm ch}$ (fm) & $\beta_{20}$&$\Delta S_\Lambda$ (MeV) &$\Delta r_{\rm ch}$ (fm)\\
		\hline
		{} & \multicolumn{3}{c}{$^{10}_\Lambda$Be}                       &  \multicolumn{3}{c}{$^{10}_\Lambda$B} \\ 
		\cmidrule(lrr){2-4}  \cmidrule(lrr){5-7}           
		PK1-Y1    & 8.138  & 2.361 & $-$0.270 & 8.110 & 2.541 & $-$0.276 & 0.028 & $-$0.180 \\
		NLSH-A    & 8.096  & 2.359 & $-$0.264 & 8.037 & 2.530 & $-$0.270 & 0.059 & $-$0.171 \\
		DD-ME2-Y1 & 8.237  & 2.389 & $-$0.266 & 8.142 & 2.558 & $-$0.273 & 0.095 & $-$0.169 \\
		DD-ME2-Y2 & 7.992  & 2.397 & $-$0.270 & 7.910 & 2.566 & $-$0.277 & 0.082 & $-$0.169 \\
		DD-ME2-Y3 & 7.631  & 2.416 & $-$0.280 & 7.585 & 2.587 & $-$0.286 & 0.046 & $-$0.171 \\
		PKDD-Y1   & 8.426  & 2.360 & $-$0.268 & 8.345 & 2.536 & $-$0.274 & 0.081 & $-$0.176 \\
		PKDD-Y2   &8.442   & 2.369 & $-$0.271 & 8.366 & 2.544 & $-$0.276 & 0.076 & $-$0.175 \\
		PKDD-Y3   &7.416   &2.390  & $-$0.281 & 7.377 & 2.566 & $-$0.287 & 0.039 & $-$0.176 \\
		Exp.      &9.11(22)~\cite{Cantwell1974_NPA236-445}
		&       &          &8.89(12)~\cite{Davis1986_CP27-91}
		&       &          & 0.22  &          \\
		&  8.60(23)~\cite{Gogami2016_PRC93-034314} 
		&       &          & 8.64(10)~\cite{Gogami2016_PRC93-034314}
		&       &          &$-0.04$&        \\                        
		{}        & \multicolumn{3}{c}{$^{12}_\Lambda$B}      &  \multicolumn{3}{c}{$^{12}_\Lambda$C}   \\ 
		\cmidrule(lrr){2-4}  \cmidrule(lrr){5-7}                            
		PK1-Y1    &9.968   & 2.401 & $-$0.231 &9.953  &2.517  & $-$0.238 & 0.015  & $-$0.116 \\
		NLSH-A    &10.067  & 2.409 & $-$0.223 & 10.072& 2.518 & $-$0.229 &$-$0.005& $-$0.109 \\
		DD-ME2-Y1 &10.466  & 2.428 & $-$0.238 & 10.391& 2.527 & $-$0.244 & 0.075  & $-$0.099 \\
		DD-ME2-Y2 &10.141  & 2.435 & $-$0.241 & 10.092&  2.534& $-$0.248 & 0.049  & $-$0.099 \\
		DD-ME2-Y3 &9.616   & 2.453 & $-$0.251 &9.622  & 2.554 & $-$0.258 &$-$0.006& $-$0.101 \\
		PKDD-Y1   &10.534  & 2.411 & $-$0.253 & 10.500& 2.521 & $-$0.259 & 0.034  & $-$0.110 \\
		PKDD-Y2   &10.526  & 2.419 & $-$0.254 & 10.502& 2.528 & $-$0.259 & 0.024  & $-$0.109 \\
		PKDD-Y3   &9.319   & 2.440 & $-$0.265 & 9.323 & 2.549 & $-$0.271 &$-$0.004& $-$0.109 \\
		Exp.      & 11.529(25)~\cite{Tang2014_PRC90-034320}
		&       &          & 11.30(19)~\cite{Gogami2016_PRC93-034314}
		&       &          & 0.23   &         \\
		{}       &\multicolumn{3}{c}{$^{16}_{~\Lambda}$N}       &\multicolumn{3}{c}{$^{16}_{~\Lambda}$O}        \\
		\cmidrule(lrr){2-4}  \cmidrule(lrr){5-7}
		PK1-Y1   & 12.140 & 2.608  & 0.000    & 12.147& 2.703 & 0.000    & $-0.007$& $-0.095$ \\
		NLSH-A   & 12.438 & 2.618  & 0.000    &12.449 & 2.709 & 0.000    & $-0.011$& $-0.091$ \\
		DD-ME2-Y1& 13.043 & 2.612  & 0.000    & 12.976& 2.692 & 0.000    & 0.067   & $-0.080$ \\
		DD-ME2-Y2& 12.739 & 2.617  & 0.000    & 12.682& 2.697 & 0.000    & 0.057   & $-0.080$ \\
		DD-ME2-Y3& 12.221 & 2.631  & 0.000    & 12.189& 2.711 & 0.000    & 0.032   & $-0.080$ \\
		PKDD-Y1  & 13.297 & 2.588  & 0.000    & 13.246& 2.677 & 0.000    & 0.051   & $-0.089$ \\
		PKDD-Y2  & 13.268 & 2.595  & 0.000    & 13.230& 2.684 & 0.000    & 0.038   & $-0.089$ \\
		PKDD-Y3  & 11.942 & 2.611  & 0.000    & 11.929& 2.700 & 0.000    & 0.013   & $-0.089$ \\
		Exp.     & 13.76(16)~\cite{Cusanno2009_PRL103-202501} 
		&        &          & 12.42(5)~\cite{Hashimoto2006_PPNP57-564} 
		&       &          & 1.34    &          \\
		{}       & \multicolumn{3}{c}{$^{40}_\Lambda$K}      &  \multicolumn{3}{c}{$^{40}_\Lambda$Ca}   \\
		\cmidrule(lrr){2-4}  \cmidrule(lrr){5-7}		
		PK1-Y1   &18.509  & 3.401  & $-$0.035 &18.497 &3.441 & $-$0.036  & 0.012   & $-$0.040 \\
		NLSH-A   &18.864  & 3.413  & $-$0.038 &18.868 & 3.453& $-$0.037  &$-$0.004 & $-$0.040 \\
		DD-ME2-Y1&19.773  & 3.403  & $-$0.025 &19.710 &3.437  &$-$0.026  &0.063    & $-$0.034 \\
		DD-ME2-Y2&19.548  & 3.406  & $-$0.025 &19.493 & 3.440 & $-$0.026 & 0.055   & $-$0.034 \\
		DD-ME2-Y3&19.318  & 3.414  & $-$0.026 &19.283 &3.448  & $-$0.026 & 0.035   & $-$0.034 \\
		PKDD-Y1  &19.990  & 3.383  & $-$0.028 &19.933 &3.420  & $-$0.030 & 0.057   & $-$0.037 \\
		PKDD-Y2  &20.170  & 3.387  & $-$0.029 &20.124 & 3.425 & $-$0.030 & 0.046   & $-$0.038 \\
		PKDD-Y3  &19.047  & 3.397  & $-$0.030 &19.025 & 3.434 & $-$0.031 & 0.022   & $-$0.037 \\
		Exp.     &        &        &          &18.7(11)~\cite{Usmani1999_PRC60-055215}\\
		\hline         
		\hline	
	\end{tabular}
	\label{Tab:gs_properties_hypernuclei}
\end{table*}

\begin{figure}[htbp]
	\centering
	\includegraphics[width=0.45\textwidth]{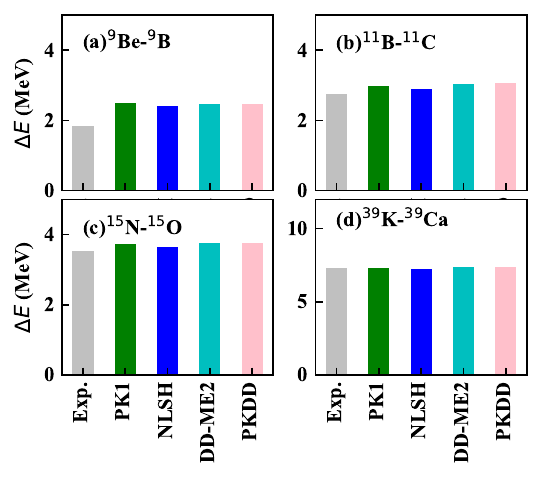}
	\caption{(Color online) The binding energy differences of mirror nuclei (a) $^9$Be-$^9$B, (b) $^{11}$B-$^{11}$C, (c) $^{15}$N-$^{15}$O, and (d) $^{39}$K-$^{39}$Ca. For deformed nuclei, rotational energy correction is explicitly incorporated into the total energy calculations. The corresponding experimental values are taken from Ref.~\cite{Wang2021_ChinPhysC45-030003}.}
	\label{Fig:DE_RHB} 
\end{figure}

\section{Results and discussion}\label{sec:results}

This study systematically investigates four mirror hypernuclear pairs: 
$^{10}_\Lambda$Be-$^{10}_\Lambda$B, $^{12}_\Lambda$B-$^{12}_\Lambda$C, $^{16}_\Lambda$N-$^{16}_\Lambda$O, and $^{40}_\Lambda$K-$^{40}_\Lambda$Ca. 
We first perform calculations for the ground-state properties of their corresponding mirror nuclear cores (
$^{9}$Be-$^{9}$B, $^{11}$B-$^{11}$C, $^{15}$N-$^{15}$O, and $^{39}$K-$^{39}$Ca) using four density functionals:  PK1~\cite{Long2004_PRC69-034319}, NLSH~\cite{Sharma1993_PLB312-377}, DD-ME2~\cite{Lalazissis2005_PRC71-024312}, and PKDD~\cite{Long2004_PRC69-034319}. The calculated nuclear properties are presented in Table~\ref{tab:core-nuclei}.

Our calculations reveal oblate deformations ($\beta_{20}\approx -0.35$) for both $^{9}$Be and $^{9}$B. The PK1, DD-ME2, and PKDD functionals underestimate binding energies by approximately 3 MeV, while NLSH produces overestimations. For $^{9}$Be, the calculated charge radii are smaller than experimental value~\cite{Angeli2013_ADNDT99-69}.
The $^{11}$B-$^{11}$C pair maintains oblate shapes across all density functionals, with both binding energies and charge radii exceeding experimental measurements. The 
$^{15}$N and $^{15}$O systems exhibit spherical symmetry, displaying calculated binding energies and radii in close agreement with experimental data. Similarly, near spherical configurations emerge for the 
$^{39}$K and $^{39}$Ca pair, whose calculated energies and radii show comparable consistency with measurements.
The energy differences, defined as $\Delta E= E(N,Z)-E(Z,N)$, for the four mirror nuclear pairs are presented in Table~\ref{tab:core-nuclei} and Fig.~\ref{Fig:DE_RHB}. The calculated 
$\Delta E$ values exhibit positive magnitudes with enhancement as mass number increases. Notably, while binding energies are underestimated by approximately 3 MeV for light mirror pairs, all selected effective interactions yield $\Delta E$ values that agree with experimental data within 1 MeV uncertainty.

\begin{figure*}[htbp]
	\centering
	\includegraphics[width=1.0\textwidth]{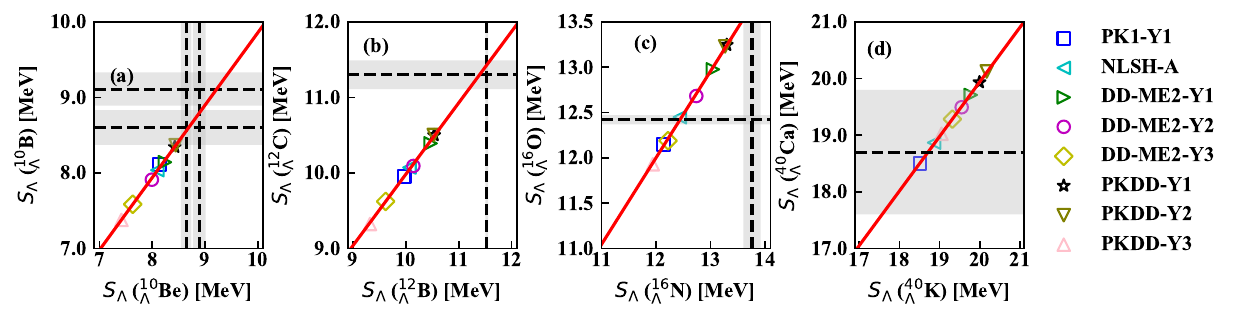}
	\caption{(Color online) The $\Lambda$ separation energy of $^A_\Lambda (Z+1)$ as a function of the $\Lambda$ separation energy of its mirror hypernucleus $^A_\Lambda Z$.
		The results are calculated using the PK1-Y1~\cite{Wang2013_CTP60-479}, NLSH-A~\cite{Mares1994_PRC49-2472}, DD-ME2-Y$i$ ($i=1,~2,~3$), and PKDD-Y$i$ ($i=1,~2,~3$)~\cite{Rong2021_PRC104-054321} effective interactions. The black dotted lines represent experimental data, while the grey shaded areas denote the corresponding uncertainties. The red curves are obtained by fitting to the eight theoretical points.}
	\label{Fig:SL}
\end{figure*}

\begin{figure*}[htbp]
	\centering
	\includegraphics[width=0.9\textwidth]{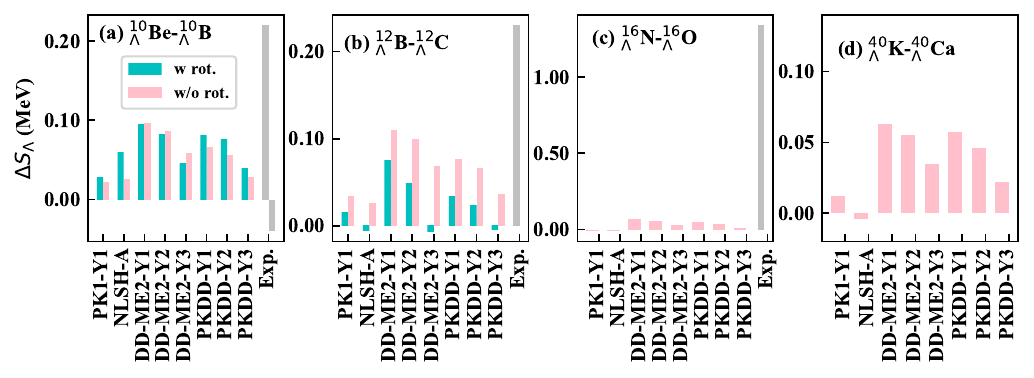}
	\caption{(Color online) The differences in $\Lambda$ separation energies of four mirror hypernuclear pairs (a) $^{10}_\Lambda$Be-$^{10}_\Lambda$B, (b) $^{12}_\Lambda$B-$^{12}_\Lambda$C, (c) $^{16}_\Lambda$N-$^{16}_\Lambda$O, and (d) $^{40}_\Lambda$K-$^{40}_\Lambda$Ca calculated using the PK1-Y1~\cite{Wang2013_CTP60-479}, NLSH-A~\cite{Mares1994_PRC49-2472}, DD-ME2-Y$i$ ($i=1,~2,~3$), and PKDD-Y$i$ ($i=1,~2,~3$)~\cite{Rong2021_PRC104-054321} effective interactions. The experimental values are shown by grey strips. For $^{10}$Be and $^{10}$B, two different measurements are presented. }
	\label{Fig:DSL}
\end{figure*}

\begin{figure*}[htbp]
	\centering
	\includegraphics[width=1.0\textwidth]{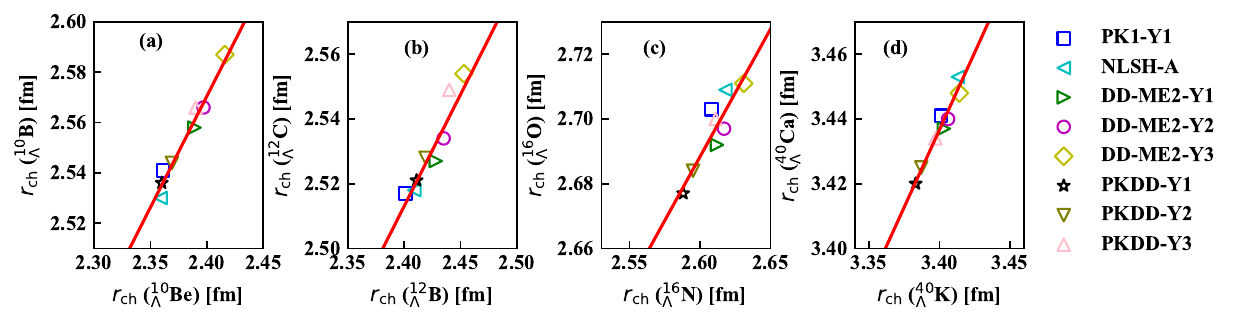}
	\caption{(Color online) Same as Fig.~\ref{Fig:SL} but for charge radii. }
	\label{Fig:Rch}
\end{figure*}

Then we investigate the ground-state properties of the selected mirror hypernuclei. The results obtained using the PK1-Y1~\cite{Wang2013_CTP60-479}, NLSH-A~\cite{Mares1994_PRC49-2472}, DD-ME2-Y$i$ ($i=1,~2,~3$), and PKDD-Y$i$ ($i=1,~2,~3$)~\cite{Rong2021_PRC104-054321} effective interactions are listed in Table~\ref{tab:hypernuclei-gs}.
The $\Lambda$ separation energy is defined as $S_\Lambda= E( ^A_\Lambda Z)-E( ^{A-1} Z)$.
For $A=10$ mirror hypernuclei, 
the experimental $S_\Lambda$ values are 9.11$\pm$0.22 MeV for $^{10}_\Lambda$Be from emulsion data~\cite{Cantwell1974_NPA236-445} and 8.60$\pm$0.07$\pm$0.16 MeV from ($e,e'K^+$) measurements~\cite{Gogami2016_PRC93-034314}, and those for $^{10}_\Lambda$B are 8.89$\pm$0.12 MeV from emulsion data~\cite{Davis1986_CP27-91} and 8.64$\pm$0.10$\pm$0.16 MeV from ($\pi^+,K^+$) experiments~\cite{Gogami2016_PRC93-034314}.
Our calculations reveal oblate shapes for both these two hypernuclei, characterized by reduced $\beta_{20}$ deformation parameters compared to their core nuclei, consistent with the $s$-orbital induced shrinkage effect for $\Lambda$ hyperon~\cite{Hiyama1999_PRC59-2351,Tanida2001_PRL86-1982,Chen2021_SciChinaPMA64-282011}. However, the calculated $S_\Lambda$ values underestimate the experimental data.
Similar oblate configurations emerge for the $A=12$ mirror pair $^{12}_\Lambda$B and $^{12}_\Lambda$C, though the calculated $S_\Lambda$ values remain below the experimental benchmarks: 11.529$\pm$0.025 MeV for $^{12}_\Lambda$B~\cite{Tang2014_PRC90-034320} and 11.30$\pm$0.19 MeV for $^{12}_\Lambda$C~\cite{Gogami2016_PRC93-034314}. 
For $A=16$ systems $^{16}_\Lambda$N and $^{16}_\Lambda$O, spherical shapes are predicted with $S_\Lambda$ values in closer agreement with data~\cite{Cusanno2009_PRL103-202501,Hashimoto2006_PPNP57-564}. Following the trend in Ref.~\cite{Rong2021_PRC104-054321}, we observe that reduced $g_{\sigma \Lambda}$ coupling constants enhance $S_\Lambda$ values when employing identical $NN$ interactions.
The $A=40$ mirror pair $^{40}_\Lambda$K and $^{40}_\Lambda$Ca exhibit near spherical configurations with negative $\beta_{20}$ deformations. While the calculated $S_\Lambda$ for $^{40}_\Lambda$Ca shows slight overestimation, most results remain within experimental uncertainty. The predicted $S_\Lambda$ values for $^{40}_\Lambda$K generally exceed those of its mirror partner across all considered interactions except NLSH-A.

To achieve the correlation of the $\Lambda$ separation energies between mirror pairs, the $S_\Lambda$ of the $^{A}_\Lambda Z$ hypernucleus is plotted as a function of $S_\Lambda$ for its mirror hypernucleus in Fig.~\ref{tab:hypernuclei-gs}. 
One can find from this figure that when calculated with different effective interactions, the results are strong linear correlations across all four mirror hypernuclear pairs. 
The data are fitted with the linear relation $S_{\Lambda}[^A_\Lambda (Z+1)]=k\cdot S_{\Lambda} [^A_\Lambda Z] +b$ with the corresponding slopes $k$ and intercepts $b$ tabulated in Table~\ref{tab:lines-property}. 
In the table, the slopes are slightly less than unity while the intercepts remain positive, implying that $S_\Lambda$ values for hypernuclei with proton number $Z+1$ exceed those of their mirror counterparts with proton number $Z$. 
Given the energy $E( ^{A-1}Z)>E(^{A-1}(Z+1))$ for normal nuclei, this demonstrates that the $\Lambda$ hyperon amplifies the energy difference between mirror nuclei--specifically, the isospin effect becomes enhanced through $\Lambda$ injection into nuclear systems. 

\begin{table}[htb!]
	\caption{The slope $k$ and intercept $b$ of the linear function $f[^A_\Lambda (Z+1)]=k\cdot f [^A_\Lambda Z] +b$ with mass numbers $A=10,12, 16$, and 40, where $f$ represents either the $\Lambda$ separation energy $S_{\Lambda}$ or charge radius $r_{\rm ch}$.}\label{tab:lines-property} 
	\doublerulesep 0.1pt \tabcolsep 12pt
	\begin{tabular}{ccccc}
		\hline
		\hline
		&  \multicolumn{2}{c}{$S_\Lambda$}  & \multicolumn{2}{c}{$r_{\rm ch}$}  \\
		\cmidrule(lrr){2-3} \cmidrule(lrr){4-5}
		$A$   &  $k$  &  $b$  & $k$       &  $b$\\
		\hline
		10  &  0.961  & 0.254 &  0.883    &  0.452  \\
		12  &  0.944  & 0.540 &  0.690    &  0.857 \\
		16  &  0.957  & 0.520 &  0.789    &  0.638  \\
		40  &  0.965  & 0.637 &  0.956    &  0.187  \\
		\hline         
		\hline	
	\end{tabular}
\end{table}

We quantify the CSB effect through the $\Lambda$ separation energy difference defined as $\Delta S_\Lambda=S_\Lambda[^A_\Lambda Z]-S_\Lambda[^A_\Lambda (Z+1)]$ for each interaction.
Fig.~\ref{Fig:DSL} presents the $\Delta S_\Lambda$ values calculatd with the eight selected effective interactions.
The calculated values are below 1 MeV across all cases.
For a fixed $NN$ interaction, an enhanced $g_{\sigma\Lambda}$
coupling strength decreases $\Delta S_\Lambda$.
For instance, when employing the DD-ME2 parametrization for $NN$ interaction, the $\Delta S_\Lambda$
magnitudes obtained with DD-ME2-Y1 ($g_{\sigma\Lambda}=0.366g_{\sigma N}$) exceed those from DD-ME2-Y2 (
$g_{\sigma\Lambda}=0.417g_{\sigma N}$) and DD-ME2-Y3 ($g_{\sigma\Lambda}=0.577g_{\sigma N}$).
Given the SU(3) flavor symmetry prediction of 
$g_{\sigma\Lambda}/g_{\sigma N}\approx 0.667$, our findings demonstrate a positive correlation between the magnitude of CSB effects and the degree of SU(3) symmetry breaking with $g_{\omega\Lambda}/g_{\omega N}$ deviates from 2/3 when manifested in $\Delta S_\Lambda$.  

Based on the deformation characteristics of $A=10$ and $A=12$ hypernuclei, we present the results without rotational corrections for these mirror pairs for comparative analysis. 
Figure~\ref{Fig:DSL}(a) reveals that the rotational energy correction enhances the 
$\Delta S_\Lambda$ values between $^{10}_\Lambda$Be and $^{10}_\Lambda$B calculated with PK1-Y1, NLSH-A, and PKDD-Y$i$ ($i=1,2,3$) parameterizations, whereas it reduces the values when using DD-ME2-Y2 and DD-ME2-Y3.
The rotational correction effect becomes more pronounced in the $A=12$ hypernuclear system, where the energy differences decrease and even exhibit negative values with enhanced $\sigma-\Lambda$ coupling strengths.
Recent work by Sun et al. has proposed a phenomenological CSB $\Lambda N$ interaction within the RMF formalism to account for the experimental binding energy disparity in the $^{12}_\Lambda$B-$^{12}_\Lambda$C mirror pair~\cite{Sun2025_PLB865-139460}. However, spherical symmetry is restricted in their calculations. Our analysis suggests that constraints derived from $\Lambda$ separation energy differences in $A=10$ and $A=12$ hypernuclear pairs must explicitly incorporate rotational correction effects, i.e., the rotational correction need to be considered before fitting the CSB term.

\begin{figure}[htbp]
	\centering
	\includegraphics[width=0.45\textwidth]{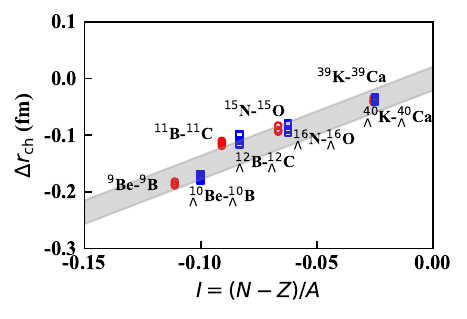}
	\caption{(Color online) The differences in charge radii of the four mirror hypernuclear pairs
		as a function of the isospin asymmetry parameter $I$. The gray band corresponding to the empirical relationship $\Delta r_{\rm ch}=1.574I\pm 0.021$ reported in Ref.~\cite{Novario2023_PRL130-032501}.}
	\label{Fig:Drch_RHB} 
\end{figure}

The charge radius $r_{\rm ch}$ exhibits modifications upon 
$\Lambda$ hyperon injection. As shown in Table~\ref{tab:core-nuclei} and Table~\ref{Tab:gs_properties_hypernuclei}, most hypernuclei demonstrate reduced $r_{\rm ch}$ values relative to their core nuclei, indicating 
$s-$orbital induced nuclear shrinkage~\cite{Hiyama1999_PRC59-2351,Takahashi2001_PRL87-212502}.
Figure~\ref{Fig:Rch} displays the linear correlation between 
$r_{\rm ch}$ values of $^{A}_\Lambda Z$ hypernuclei and their mirror partners. 
Analysis of slopes in Table~\ref{tab:lines-property} reveals smaller values for charge radii than those for $\Lambda$ separation energies.
The hypernuclear charge radius difference, defined as 
$\Delta r_{\rm ch}=r_{\rm ch}[^A_\Lambda Z]-r_{\rm ch}[^A_\Lambda (Z+1)]$, maintains negative values with magnitudes comparable to those of the corresponding core nuclei differences. This preserves the characteristic charge-radius hierarchy where proton-rich systems exhibit larger radii than neutron-rich counterparts, observed in both normal nuclei and hypernuclei.
The dependence of $\Delta r_{\rm ch}$ on isospin asymmetry is plotted in Fig.~\ref{Fig:Drch_RHB}. Both normal and hypernuclear data align closely with the linear relation 
$\Delta r_{\rm ch}=1.574I\pm 0.021$ established through \textit{ab initio} calculations~\cite{Novario2023_PRL130-032501}.

\section{Summary}\label{sec:summary}

In-medium $YN$ interaction is typically constrained by the hyperon separation energies in a hypernucleus. However, in contrast to normal nuclei, within the mean-field framework, the properties of mirror hypernuclei are seldom utilized to constrain hyperon-nucleon interactions. In this study, we examine the ground-state properties of four pairs of mirror hypernuclei: $^{10}_\Lambda$Be-$^{10}_\Lambda$B, $^{12}_\Lambda$B-$^{12}_\Lambda$C,  $^{16}_\Lambda$N-$^{16}_\Lambda$O, and $^{40}_\Lambda$K-$^{40}_\Lambda$Ca. 
The theoretical framework adopted in this study involves a deformed relativistic Hartree-Bogoliubov model, which is employed to elucidate the relationships between the aforementioned hypernuclei and the effective $\Lambda N$ interactions.

The calculated results, obtained with eight effective interactions, reveal linear correlations for both separation energies and charge radii of the studied mirror hypernuclear pairs. 
The results from the difference of the $\Lambda$ separation energy indicate that the larger the charge symmetry breaking, the larger the breaking of the SU(3) flavor symmetry for the scalar coupling constant $g_{\sigma\Lambda}$. 
It is proposed that the contribution of rotational correction
energy needs to be accounted for before utilizing the $\Lambda$ separation energy differences of $A=10$ and $A=12$ pairs to impose constrains. 
Besides, measurements for $A=16$ and $A=40$ mirror hypernuclei will provide enhanced constraints on isospin-dependent $YN$ interactions.

\section*{Acnowledgements}

This work has been supported by the National Natural Science Foundation of China (12205057), the Science and Technology Plan Project of Guangxi (Guike AD23026250), the Natural Science Foundation of Guangxi (2024JJB110017, 2023GXNSFAA026016), the National Natural Science Foundation of China (12365016), the Central
Government Guides Local Scientific and Technological Development Fund Projects (Guike ZY22096024), the Natural Science Foundation of Henan Province (242300421156), and the National Natural Science Foundation of
China (U2032141). The results shown in this paper are obtained through the Guangxi Key Laboratory of Nuclear Physics and Technology High Performance Computing Platform.

	\bibliographystyle{elsarticle-num_20210104-sgzhou}
	\bibliography{ref.bib}
	
\end{document}